\documentclass[twoside]{article}
\usepackage{bourbaphy_h}
\usepackage{graphicx}
\usepackage{amsmath,amssymb,amsfonts,bm,slashed,graphics,color}
\usepackage{graphicx}
\usepackage{cite}
\usepackage{pdfsync}
\usepackage{tensor}
\usepackage{subfigure}
\usepackage[colorlinks=true]{hyperref}
\def\bee{\begin{equation}}
\def\ee{\end{equation}}
\def\bea{\begin{eqnarray}}
\def\eea{\end{eqnarray}}
\def\ba{\begin{array}} %%%%%%%
\def\ea{\end{array}}
\def\bc{\begin{center}}
\def\ec{\end{center}}

\def\ghost#1{}
\def\sl{\!\!\!/}
\def\Tr{\text{Tr}}

\def\simge{\mathrel{%
   \rlap{\raise 0.511ex \hbox{$>$}}{\lower 0.511ex \hbox{$\sim$}}}}
\def\simle{\mathrel{
   \rlap{\raise 0.511ex \hbox{$<$}}{\lower 0.511ex \hbox{$\sim$}}}}

\newcommand{\cB}{ { \cal{B} } }
\newcommand{\cG}{ { \cal{G} } }

\begin{document}

%\large
\title{Quantum Gravity and Random Tensors}
\author{R. Gurau\\
University of Heidelberg\\
V. Rivasseau\\
Universit\'e Paris-Saclay
%, CNRS/IN2P3\\ IJCLab, 91405 Orsay, France
}

\maketitle

\begin{abstract}

%\vspace{1mm}

Random tensors are the natural generalization of random matrices to higher order objects. They provide generating functions for random geometries and, assuming some familiarity with random matrix theory and quantum field theory, we discuss in the first part of this note the applications of such models to quantum gravity. In a second part we review tensor field theories, that is standard field theories in $\mathbb{R}^d$ but with tensor fields, which lead to a new family of large $N$ conformal field theories relevant for the study of the $AdS/CFT$ correspondence.
\end{abstract}

\tableofcontents

\section{Introduction}

Random tensors \cite{RG}, like random matrices \cite{Wigner}, originated in theoretical physics.  The central goal of the 1970's theoretical physics was the quantization of elementary particles like quarks and gluons. Matrix models, and in particular their large $N$ limit, turned out to be of relevance for the quantization of the theory of strong interactions \cite{tHooft:1973alw}. Shortly after, random  matrices where understood to provide the correct framework for the study of two dimensional random (quantum) surfaces \cite{Brezin:1977sv,Ambjorn:1985az,DiFrancesco:1993cyw} and of statistical models on such surfaces \cite{Kazakov:1986hu,Brezin:1989db}.
This line of work culminated in the formulation of the KPZ relation \cite{Knizhnik:1988ak} linking the critical exponents of a statistical model in random geometry to the critical exponents of the same model in fixed geometry.
Together with the counter intuitive realization that critical exponents are easier to compute for the random geometry versions, this provided a new approach to the study of statistical models and conformal field theory in two dimensions.

As the world sheet of a one dimensional string is two dimensional, the theory of quantum surfaces encoded in random matrices has been extensively explored since the 90's in the context of string theory \cite{Brezin:1990rb,Douglas:1989ve,Gross:1989vs}, by far the most mature approach to quantum gravity to this date. The continuum limit of random matrices is understood today in terms of Liouville quantum gravity \cite{Duplantier:2014daa,DuplKPZ} and the matter content in terms of Schramm Loewner Evolution \cite{DuplaSLE}, see the accompanying contribution by 
J.~Miller. The limit of random discretized surfaces,  called the Brownian sphere \cite{LeGall}, is by now a well understood mathematical object, see the accompanying contribution by J.~F.~Le Gall.

Random matrices have a rich set of applications going well beyond random surfaces.
In quantum chaos, the Bohigas--Giannoni--Schmit conjecture asserts that the spectral statistics of quantum systems whose classical counterpart exhibits chaotic behaviour is described by random matrix theory \cite{BohGiaSch}.
In number theory, the distribution of zeros of the Riemann zeta function is modeled by the distribution of eigenvalues of certain random matrices, as pointed out by Montgomery \cite{Mon} and Dyson. In free probability theory, pioneered by Voiculescu \cite{Voi} in an algebraic context,  the restriction to non crossing partitions in the definition of free cumulants is modeled on the $N\to \infty$ planar limit of random matrices and, unsurprisingly, the limit distribution in the free central limit theorem reproduces the Wigner's semi-circle law. 
In computational theory, one often considers random operators that generate families of computational models, in the hope to improve performance \cite{Soi}; in the theory of optimal control one uses random matrices in the state equation leading to stochastic control \cite{Cho}. And the list goes on.
 
While this is by no means an exhaustive list of applications of random matrices, or relevant references on them, it illustrates their far reaching successes as versatile tools in physics and mathematics. 

\paragraph{Random tensors.}
Originally proposed in the '90s, random tensor models \cite{Ambjorn:1990ge,Gross:1991hx,Sasakura:1990fs} aimed to
generalize to higher dimensions the connection between random matrices and random surfaces. The idea was to construct generating functions for random higher dimensional spaces using tensors of order three or more. However, trying to replicate the success of two dimensional theories in higher dimensions is a daunting challenge, not in the least because a continuous theory of quantum gravity, generalizing for instance Liouville quantum gravity, is not available in three or more dimensions. As one does not have a good description of continuum quantum gravity, one does not really know what to aim for in the discrete setting: besides vague requirements like ``one would like to find a nice smooth Lorentzian manifold with an Einstein-Hilbert effective action in some limit'', anything could in principle go. 

It follows that in higher dimensions one is naturally pushed to survey the landscape of possible theories of discrete spaces and identify universal features which are reproduced in all (or most) of them. One such feature is that, if generated by random tensors, the discrete higher dimensional geometries are naturally organized in powers of $1/N$, with $N$ the dimension of the Hilbert space on which the tensors act. While by today this is a well established fact, due to some technical subtleties
\cite{Gurau:2009tw}, establishing this $1/N$ expansion for tensors requires a bit more work than in the case of matrices 
\cite{Gurau:2010nd,Gurau:2010ba,Gurau:2011aq,Gurau:2011xq,Gurau:2013pca,Dartois:2013he,Gurau:2017qya,Ferrari:2017jgw,Klebanov:2017nlk,
Benedetti:2017qxl,Bonzom:2019moj,Benedetti:2020iyz,Carrozza:2020eaz,Carrozza:2021qos}.
Contrary to matrices, for which the corresponding $1/N$ expansion is indexed by the genus and is therefore an expansion in topologies, for higher order tensors the $1/N$ expansion is indexed by a non negative half-integer, called 
the \emph{degree}\footnote{Sometimes referred to as the Gurau degree in the literature.}, which is not a topological invariant. Be that as it may, one at least has at ones disposal a starting point to study the ensuing random space: the \emph{melonic} \cite{Bonzom:2011zz} large $N$ limit, which is universal. 

Indeed, the world of random tensors has a surprisingly simple access door, the family of melonic graphs \cite{Bonzom:2011zz,BonGurRiv} which is encountered at leading order in the $1/N$ expansion. Despite its simplicity (see Figure~\ref{fig:melonic} for some examples), the family of melonic graphs,
gives rise to surprisingly complex behaviour. 

\paragraph{Generalized Kontsevich models.}
Based on the $1/N$ expansion of tensors, the tensor track 
\cite{tensortrack1,tensortrack2,tensortrack3,tensortrack4,tensortrack5,tensortrack6} program aims to formulate a reasonable theory of random geometry and ultimately provide a version of discretized quantum gravity in dimensions greater than two. The cornerstone of this program is the combination of random tensor models with a renormalization group flow in an abstract discrete space. This approach is closely related to dynamical triangulations, either in their Euclidean version \cite{EDTAmbjorn}, or in their more constraint ``causal'' one \cite{LolAmbJur,Lol}. It is a so called ``background independent'' approach to quantum gravity, which does not presuppose the existence of some embedding space. On the contrary, it aims to generate the space ab initio, out of nothing. 

Tensor models with invariant interactions and a non invariant propagator have been called tensor field theories. Such models are closely related to the Kontsevich model \cite{Kontsevich:1992ti,GroWul}, and will be refereed to below as generalized Kontsevich models: we reserve the name tensor field theories for field theories with tensor fields living in a flat Euclidean $\mathbb{R}^d$ space, which will be discussed separately.
The renormalizable Kontsevich like models studied so far divide into two main categories, those with \cite{Oriti:2016acw} and those without \cite{Gel1} additional group theoretical data. The combinatorial models 
with no group theoretical data are better suited for the study of renormalization \cite{BenGeloun:2011rc,BenGeloun:2012pu,
Delepouve:2014hfa,Rivasseau:2017xbk,BenGeloun:2016tmc,
BenGeloun:2017xbd,BenGeloun:2018ekd,Eichhorn:2018ylk, 
Eichhorn:2019hsa}. The ones with additional data could be 
better suited for the study of quantum spaces 
\cite{CarOriRiv,Carro,Lahoche:2015ola,Baloitcha:2020lha,Finocchiaro:2020fhl,
Marchetti:2022igl,Goeller:2022ywz,Marchetti:2022nrf}, as the additional data can encode additional geometric information (e.g. a discretized holonomy) about the underlying space. For an up-to-date review of distinct approaches to quantum gravity including the two discussed here, see \cite{Boer}.

Results obtained so far on  Kontsevich like models include for instance renormalization theorems at all orders, constructive field theory results,
computations of beta functions at one and two loops or studies of their functional renormalization group flow.  One aim in this context would be to solve such theories in closed form at the leading melonic order using Ward identities and Schwinger-Dyson equations, as it has been achieved in the closely related Grosse-Wulkenhaar model \cite{GroWul}. Some preliminary results in this direction on toy models have been obtained \cite{Pascalie:2019yxd}.
Tensor interactions with ``derivative operators'', hence with non invariant interactions, have been introduced and investigated \cite{Gel10,GelRei2}, enlarging maximally the space of theories one can explore. 

Of course, like any serious approach to higher dimensional quantum gravity, the tensor track also has its limitations. Most importantly, the precise geometries one obtains are not yet well understood. While the spaces selected at leading order (one could call them ``melonic spaces'') are topologically spheres, the metric assigned to a such a topological space is model dependent. In the simplest models, in which both the interactions and the propagator are invariant, the natural metric assignment is to associate an equilateral triangulation to the topological triangulation represented by a graph \cite{RG}. With this metric assignment, the metric structure of melons is that of a branched polymer \cite{Gurau:2013cbh}, a space which, despite having the right topological dimension, looks like a tree for all intents and purposes. This seems to persist in a double scaling limit \cite{GurSch,DarGurRiv}.
The richer Kontsevich like models can endow the melonic triangulations with a smoother geometry, but a systematic study of the ensuing random metric space has not so far been performed.

\paragraph{Tensor field theories.}
More recently, quantum field theories for tensor fields have received increased attention. Such theories make no claims at background independence: they are just usual quantum field theories in an existing space (or space time). Nonetheless, they are still relevant for the problem of quantum gravity as the provide non trivial examples of conformal field theories (CFTs) and can thus be used to study interesting quantum gravity models via the $AdS/CFT$ correspondence.  

In 2015 the Sachdev-Ye-Kitaev (SYK) model \cite{SacYe,Kit, Kit1,MalSta2016}
has been understood to provide a simple nearly conformal dual to a nearly $AdS_2$ black hole \cite{MalStaYan} and in particular to display ``maximally'' chaotic quantum behavior \cite{MalStaShe}. This fascinating remark generated a significant amount of progress on the study of holography, see e.g. \cite{Rosenhaus:2018dtp} and subsequent work.
Shortly after, it has been remarked that the melonic large $N$-limit,
which is due to a disorder average in the SYK model, is achieved naturally in a tensor quantum mechanical model \cite{witten2019syk,Gurau:2016lzk,klebanov2017uncolored,Klebanov:2018nfp,Pakrouski:2018jcc,Klebanov:2019jup,Gaitan:2020zbm}, with no need for disorder. Tensor quantum mechanics displays a genuine renormalization group flow, and one can study its infrared conformal limit with standard techniques.

In higher dimensions \cite{Giombi:2017dtl,carrozza2016n,Bulycheva:2017ilt,Giombi:2018qgp, Choudhury:2017tax,Benedetti:2018goh,Klebanov:2018fzb,Kim:2019upg,
DeMelloKoch:2019tmo,Klebanov:2020kck,Peng:2016mxj,Choudhury:2021nal,Jepsen:2023pzm} tensor field theories provide a new class of conformal field theories, dubbed melonic CFTs \cite{Benedetti:2019eyl,Benedetti:2018ghn,Gurau:2019qag,Benedetti:2019rja,Benedetti:2019sop,Benedetti:2019ikb,Benedetti:2020iku,Harribey:2021xgh,Benedetti:2021wzt}. They are examples of large $N$ CFTs that are at the same time non trivial and amenable to analytic computations. This is somewhat surprising:  although, as algebraic objects, higher order tensors are more complicated than both vectors and matrices, the large $N$ CFTs obtained in tensor field theories are simpler than the ones of matrix field theories. This is due to the fact that the melonic family, while richer than the family of cacti (or snail \cite{Klebanov:2018fzb}) diagrams which dominate at large N in vector models, are much simpler than the family of all the planar diagrams one encounters at leading order in the matrix case. 

Tensor field theories are also interesting at their upper critical dimension and it turns out that (in Euclidean signature) one finds models that display asymptotic freedom while maintaining boundedness from below of the real part of the action \cite{Berges:2023rqa}.

\paragraph{Plan of this paper.}
In the next two sections we will review in more detail the two major applications of random tensors to physics. We will first discuss the relation between random tensors, random geometries and Kontsevich like models. In the second part,  we will review tensor quantum mechanical models and tensor field theories.

We stress that both these lines of inquiry are very dynamical research fields and we expect that much more is to come from each of them. 
Random tensors are expected to play an increasingly important role in the future, not only in core domains, like the study of field theories and random geometries \cite{DelRiv}, but also in the exploration of new domains like machine learning \cite{RicMon,OueTamRivI,tensortrack7}.

\section{Random tensors and Kontsevich like models}

The simplest random matrix distributions are the three Gaussian ensembles: orthogonal, unitary and symplectic. They are often denoted by their Dyson index, $\beta  = 1$ for the orthogonal case, $\beta  = 2$ for the unitary case, and $\beta  = 4$ for the symplectic one, which counts the number of real components per matrix element. The ensembles with non integer $\beta$ are equally fascinating, although they lack a formulation in terms of random matrices.

The Gaussian Unitary Ensemble (GUE) is the probability measure:
\begin{equation}
d\mu(M) = d M  \; e^{- \frac{N}{2}\Tr(M^2) }  \;, \quad
d M= K \prod_{k,l} d M_{kl}  \;,
\end{equation}
where $M$ is a Hermitian $N\times N$ matrix and $K$ is a constant chosen such that $\mu$ is normalized. 
This measure is invariant under the conjugation of $M$ with a unitary $N\times N$ matrix, $M\to UMU^\dagger$, and one is interested in the expectations of invariant observables. 

The invariant observables are readily identified:
an algebraic basis in the space of polynomial invariants is provided by the traces of the powers of $M$. The joint expectations:
\begin{equation}
 \left\langle \Tr ( M^{p_1} )  \dots \Tr ( M^{p_n} )
 \right\rangle_{GUE}  =\int d M  \; e^{- \frac{N}{2}\Tr (M^2) } \; \Tr  ( M^{p_1}  ) \dots \Tr ( M^{p_n}  ) \;, 
\end{equation}
are computed using Wick's theorem as sums over pairings (Gaussian pairings for mathematicians, or Feynman graphs for physicists) of matrix entries, where each pair $( M_{ij}, M_{kl} )$ is replaced by a ``propagator'':
\begin{equation}
    \int d M  \; e^{- \frac{N}{2}\Tr ( M^2 ) } \;  M_{ij} M_{kl}  = 
     \frac{1}{N} \delta_{il}\delta_{jk}  \;.
\end{equation}
Each pairing contributes a $N$-dependent weight (also know as Feynman amplitude) to the correlation and any expectation displays a leading large $N$ behaviour followed by corrections in $1/N$. 

More general invariant random matrix models are obtained by perturbing the GUE with an invariant perturbation, or  interaction for physicists \cite{DiFrancesco:1993cyw}. 
We stress that the interaction is required to also be invariant under conjugation by the unitary group. A common choice is to consider ``single trace'' perturbations, that is perturbations of the form $S^{\rm int}(M) = \operatorname{Tr}(f(M))$ for some function $f$. 

One is then interested in computing the expectations of the invariant observables (note the normalization):
\begin{equation}
\begin{split}
 & \left\langle \frac{1}{N}\Tr ( M^{p_1} )   \dots \frac{1}{N} \Tr ( M^{p_k} ) \right\rangle\crcr
 & \qquad \qquad = \frac{1}{ {\cal Z} }  \int dM e^{-\frac{N}{2} \Tr (M^2) - N S^{\rm int}(M) }  \; \frac{1}{N}\Tr ( M^{p_1} ) \dots \frac{1}{N} \Tr ( M^{p_k} ) \;, 
 \end{split} 
\end{equation}
where the partition function ${\cal Z}$ is the normalization factor for the perturbed measure. The most interesting features of this probability measure are:
\begin{itemize}
    \item the Feynman graphs \cite{DiFrancesco:1993cyw} of this measure represent tessellated surfaces obtained by gluing polygonal patches. Each polygonal patch is associated to a monomial in the interaction $S^{\rm int}(M) $, and the propagators represent the gluing of the patches along edges. The invariant obserables, $\Tr(M^{p_1})$ and so on, represent boundaries, that is loops of length $p_1$ and so on.
    \item the amplitude of a graph  \cite{DiFrancesco:1993cyw} is 
     $N^{2-2b- 2g }$ where $g$ is the genus of the corresponding tessellated surface and $b$ the number of boundary loops,  per connected component. 
     As each connected component must have at least one boundary component, this weight is $N^0$ or lower and the expectations are series in $1/N$.
\end{itemize}

The matrix integral fills in all the possible bulk tessellations compatible with the boundary loops and assigns to each tessellated bulk a weight dependent on its topology.

\paragraph{Tensors and invariants.} 
The construction described in the previous paragraph generalizes to higher order tensors \cite{RG}.
A complex, order $D$ covariant tensor is a $D$-linear form on $\bigotimes_{i=1}^D\mathbb{C}^N$, that is a
hypercube of $N^D$ complex numbers transforming in the tensor product of $D$ fundamental representations of the unitary group: 
\begin{equation}
  T^U_{b^1b^2\dots b^D}  = \sum_{a^1,a^2,\dots a^D} U^{(1)}_{b^1a^1} U^{(2)}_{b^2a^2} \dots U^{(D)}_{b^Da^D}  T_{a^1 a^2 \dots a^D} \; , 
\end{equation}
with $U^{(i)}$ unitary transformations. Note that the indices of the tensors are distinguished and each of them transforms with its own unitary. 

\begin{figure}[!htb]
\centering
\includegraphics[width=10cm]{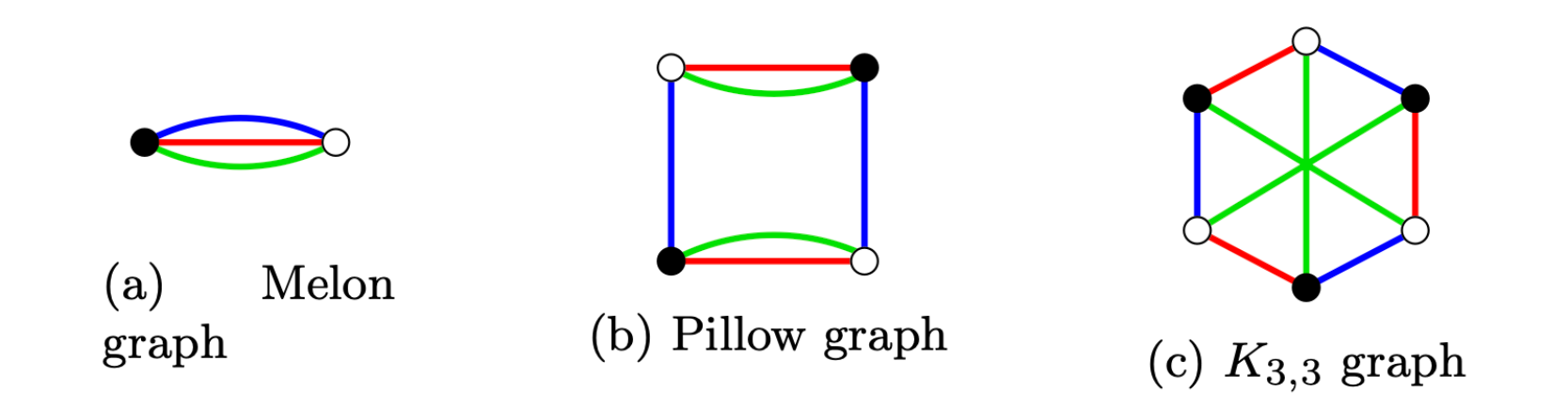}
\caption{Examples of edge 3-colored graphs. The graphs presented in the left and center panels are \emph{melonic}, while the one on the right panel is not.}
\label{fig:melonic}
\end{figure}

The first order of business is to classify the polynomial invariants one can build out of the tensor. It can be shown that any polynomial invariant is a linear combination of \emph{trace invariants} \cite{RG}. The trace invariants are represented as bipartite edge $D$-colored graphs, that is $D$-regular graphs such that every edge has a color $1,2\dots D$ and exactly one edge for each color is incident at every vertex. Some examples are depicted in Fig.~\ref{fig:melonic}. 

The invariant associated to a graph is built as follows:
\begin{itemize}
    \item for each white vertex one takes a tensor, and for each black one a complex conjugated (dual) tensor,
    \item  for each edge one has an identification of a pair of indices, respecting their position. 
\end{itemize}
In detail, denoting $V^\circ$ and $V^\bullet$ the sets of white and black vertices of a bipartite $D$-colored graph $\cB$ and $E^i$ the edge set of edges of color $i$, the trace invariant associated to $\cB$ is:
\begin{equation}
\operatorname{Tr}_{\cB}(T,\bar T ) = \sum_{\text{all indices}} \; \prod_{v\in V^\circ} T_{a^1_va^2_v \dots a^D_v}
\prod_{w\in V^\bullet} \bar T_{b^1_vb^2_v \dots b^D_v} \prod_{i=1}^D \left( \prod_{(v,w)\in E^i} \delta_{a^i_v b^i_w} \right) \;.
\end{equation}

A similar construction goes through for real tensors transforming in the tensor product of fundamental representations of the orthogonal group, if one drops the requirement of bi partition.

\paragraph{Invariant probability measures.} 
Denoting $\bar T \cdot T = \sum_{a} \bar T_{a^1a^2 \dots a^D } T_{a^1a^2 \dots a^D }$ the unique quadratic invariant on can build out of a tensor and its dual, the order $D$ tensor Gaussian unitary ensemble is the probability measure:
\begin{equation}
\begin{split}
&   d\mu(\bar T,T) =  d\bar T  dT \; e^{-\frac{N^{D-1}}{2} \bar T \cdot T } \;, \qquad
  d\bar T dT    = \prod_{a^1,\dots a^D} \frac{ d\bar T_{a^1\dots a^D} dT_{a^1\dots a^D} }{2\pi \imath } \;,
\crcr
& \int   d\bar T  dT \; e^{-\frac{N^{D-1}}{2} \bar T \cdot T }  \; T_{b^1b^2\dots b^D} \bar T_{c^1c^2 \dots c^D} =
\frac{1}{N^{D-1}} \delta_{b^1c^1} \delta_{b^2c^2} \dots \delta_{b^Dc^D} \; .
\end{split}
\end{equation}
An invariant random tensor model consists in perturbing the $D$-GUE with an invariant interaction:
\begin{equation}
    S^{\rm int}(T,\bar T)  =\sum_{ \cB } t_{\cB} \operatorname{Tr}_{\cB}(T,\bar T ) \;,
\end{equation}
where the sum runs over some set of $D$-colored graphs. We typically restrict to connected graphs $\cB$, corresponding to the single trace perturbation in random matrices. The partition function and free energy of a random tensor model are then:
\begin{equation}\label{eq:partTM}
{\cal Z}(\{t_{\cB}\};N) =  \int
d\bar T  dT \; e^{-\frac{N^{D-1}}{2} \bar T \cdot T -
 N^{D-1} S^{\rm int}( T,\bar T) } \;,\qquad 
 F(\{t_{\cB}\};N ) =\frac{1}{N^D} \log {\cal Z}(\{t_{\cB}\};N)  \;.
\end{equation}

\paragraph{Feynman graphs as triangulations.} The Feynman graphs of a random tensor model, denoted generically $\cG$, are $(D+1)$-edge colored graphs \cite{RG}. The propagator (Gaussian pairing) can be represented as a new kind of edge, say of color zero, connecting a tensor and its dual. Such graphs are Poincar\'e dual to vertex colored $D$-dimensional triangulations \cite{Lins,RG}:
\begin{itemize}
    \item for each subgraph with $D$-colors of $\cG$ one draws a triangulation vertex and assigns to it the color missing from the subgraph.
    \item two triangulation vertices are connected by a triangulation edge if their corresponding $D$-colored subgraphs share a $(D-1)$-colored subgraph.  
    \item $p+1$ vertices in the triangulation belong to a $p$ simplex if all the $D$-colored subgraphs of $\cG$ corresponding to them share the same  $(D-p)$-colored subgraph.
\end{itemize}

The observables $\Tr_{\cB}(T,\bar T)$ have only $D$ colors and represent $D-1$ dimensional boundary triangulations. The Feynman 
graphs contributing to an expectation:
\begin{equation}
\begin{split}
&    \left\langle \frac{1}{N} \Tr_{\cB_1}(T,\bar T) \dots \frac{1}{N}\Tr_{\cB_p}(T,\bar T) \right\rangle \crcr
& \qquad = \int
d\bar T  dT \; e^{-\frac{N^{D-1}}{2} \bar T \cdot T -
 N^{D-1} S^{\rm int}( T,\bar T) }  \; \frac{1}{N} \Tr_{\cB_1}(T,\bar T) \dots \frac{1}{N}\Tr_{\cB_p}(T,\bar T) \;,
 \end{split}
\end{equation}
represent $D$ dimensional bulk triangulations connecting these boundaries. 

\paragraph{The $1/N$ expansion.} The free energy as well as expectations of observables of the a random tensor model
display a $1/N$ expansion. The crux is the counting of bi colored cycles, also known as faces, in a connected $(D+1)$-colored graph \cite{Gurau:2010ba,Gurau:2011aq,Gurau:2011xq,Gurau:2019qag}:
\begin{equation}
    F(\cG) = D + \frac{D(D-1)}{4} V(\cG) - \hat \omega(\cG) \;,
\end{equation}
where $F(\cG)$ is the total number of bi colored cycles in $\cG$, $V(\cG)$ the total number of vertices of $\cG$ and $\hat \omega(\cG)$ is a non negative half integer associated to $\cG$ which furthermore respects the bounds:
\begin{equation}
    \frac{1}{D} \hat \omega(\cG) \le \hat \omega(\cG) - \hat \omega(\cG^0) \le \hat \omega(\cG) \;,
\end{equation}
where $\cG^0$ is the graph obtained from $\cG$ by erasing the edges of color $0$. 
The non negative half integer $\omega(\hat \cG)$ is called the \emph{degree} of the graph $\cG$ \cite{RG,Gurau:2013cbh} (for historical reasons it sometimes referred to as the reduced degree).

The amplitude of a connected Feynman graph 
is simply:\footnote{For a graph with $E^0(\cG)$ edges of color $0$, $K(\cG^0)$ subgraphs with colors $1,\dots D$ out of which $B(\cG)$ are boundary components of $\cG$ we have a priori 
$N^{-B(\cG)-(D-1)E^0(\cG)  +  (D-1) [ K(\cG^0) - B(\cG) ] +F(\cG) - F(\cG^0) }$ and simple manipulations lead to the result.}
\begin{equation}
N^{D  - D B(\cG) - [ \hat \omega(\cG) - \hat \omega(\cG^0)  ]  }    \;,
\end{equation}
with $B(\cG)$ the number of connected components of the boundary. This is to be compared with the familiar $N^{2 - 2 b(\cG) -2g(\cG)}$ scaling  of the matrix case.
The joint expectations of the invariants are thus series in $1/N$, and at leading order graphs with minimal $\omega(\cG)$ and the least possible number of boundary components dominate. 

The graphs with no boundary component $B(\cG)=0$ and zero
degree $\omega(\cG)=0$, which contribute to the free energy of the model, are called \emph{melonic} \cite{Bonzom:2011zz}. The melonic graphs are built recursively, starting with the ``elementary melon'' consisting in two vertices connected by $(D+1)$ edges and iteratively inserting pairs of vertices connected by $D$ edges on any of the available edges, see Fig.~\ref{fig:melonic} for some examples of meloing graphs with $2+1$ colors.

The melonic graphs are abstract triangulations of the topological sphere in any dimension $D$.\footnote{We stress that this is just a subclass of the triangulations of the sphere.} Unfortunately for the quantum gravity program, if one simply interprets these abstract triangulations as equilateral triangulations of the metric sphere, one obtains in the continuum limit the somewhat trivial continuous random tree \cite{Gurau:2013cbh} metric space, known as the branched polymer phase in the physics literature. This geometric phase seems to survive in a double scaling limit \cite{DarGurRiv,GurSch}. 

If one aims to build smoother spaces, one needs to find a way past this singular geometry phase. One way would be to try and smooth out the geometry by some sort of Ricci flow on the triangulation. Alternatively, one can consider more complicated tensor models which endow the topological triangulation with a different metric.

\paragraph{Kontsevich like models: renormalization group flow in the abstract space of indices.}
One possible future for the quantum gravity program in the discrete random geometry approach lies with considering more involved tensor models. One attempt is to impose ``causality constraints'' \cite{LolAmbJur} on the allowed higher dimensional dynamical triangulations. This approach is very successful in providing a discretized version of 
the Ho\v{r}ava–Lifshitz anisotropic theory of gravity but fails to describe Einstein-Hilbert gravity, which does not have any preferred slicing or foliation.

A mores systematic attempt to depart from the unitary invariant tensor models is to consider non unitary invariant models. However, one would like to keep some of the features of the invariant models, for instance the interpretation in terms of dual vertex colored triangulations, so it is wise to not throw the invariance completely out the window. A conservative modification is to keep an invariant interaction but alter the quadratic part. For instance one could replace the GUE free part with a non invariant covariance:
\begin{equation}
    \bar T\cdot T \to \sum_{a^1,\dots a^D} \bar T_{a^1a^2 \dots a^D } \; R^\Lambda_\mu(a^1,a^2,\dots a^D) \; T_{a^1a^2 \dots a^D }\;.
\end{equation}

One nice feature of this is that one is no longer constraint to consider tensors of finite size. One can simply start with infinite size tensors and implement
in the $R^\Lambda_\mu(a^1,a^2,\dots a^D)$ function some sort of cutoff. One can then use well known renormalization group techniques, like perturbative computations, 
\cite{BenGeloun:2011rc,BenGeloun:2016tmc,BenGeloun:2017xbd}
constructive field theory methods \cite{Delepouve:2014hfa,Rivasseau:2017xbk} or 
functional renormalization methods \cite{BenGeloun:2018ekd,Eichhorn:2018ylk,Eichhorn:2019hsa}
to study the behaviour of the partition function and correlations when varying the cutoffs. 

We stress that in these models one deals with a new kind of renormalization group flow. Contrary to the usual space time picture in which the renormalization transformation decimates the high energy degrees of freedom and rescales the system, this renormalization group flow decimates ``the components of the tensor with large indices''. As there is no notion of energy associated to such degrees of freedom, contrary to the usual case, both the cutoff and the operators (in the sense of quantum field theory) are dimensionless. It follows that one can not naively classify the operators by power counting into marginal, relevant or irrelevant and one might worry that the entire notion of renormalization of such models is nonsensical. The situation is rescued by the $1/N$ expansion of the invariant models: it is this expansion that allows one to perform a classification of the operators in terms of ``power counting'' in a dimensionless cutoff $N$ (or at least to guarantee it exists), a prerequisite of any renormalization group analysis.

Also, and perhaps even more subtle, there is \emph{no notion} of infrared behaviour in such theories as they are ultimately cutoffed at the index $0$. In particular this gives rise to finite renormalization group flows\footnote{That is couplings which exhibit a non trivial change with the scale but are neither zero nor divergent along the entire trajectory.} which do not connect renormalization group fixed points. Indeed, an ultraviolet fixed point suffices to render all the trajectories finite. The most promissing feature of this approach is that ultraviolet fixed points with a small number of relevant directions are ubiquitous \cite{BenGeloun:2011rc,BenGeloun:2016tmc,BenGeloun:2017xbd,BenGeloun:2018ekd,Eichhorn:2018ylk,Eichhorn:2019hsa}: the Kontsevich like models tend to be asymptotically safe or free, hence ultraviolet complete.

Other, more drastic, modifications of the invariant tensor models have been studied \cite{CarOriRiv,Carro,Lahoche:2015ola,Baloitcha:2020lha,Finocchiaro:2020fhl,
Marchetti:2022igl,Goeller:2022ywz,Marchetti:2022nrf} in the hope of identifying models with a better geometrical interpretation, or closer to phenomenology. The challenge in this approach is to identify some collective state, corresponding to an already formed space time, and study the perturbations around it.

\section{Tensor field theories, from the SYK model to melonic fixed points}

A second set of applications of random tensors comes under the guise of tensor field theories. These are just quantum field theories in $\mathbb{R}^d$ space, of the garden variety, but with tensor fields. Due to the melonic large $N$ limit characteristic of random tensors, tensor field theories have a very rich behaviour. 

\subsection{The SYK model and tensor quantum mechanics}

Chaos in a quantum system should be captured by out of time order correlators. For example one expects that for a quantum system in equilibrium at temperature $\beta^{-1}$ (that is in a thermal state), an out of time order four point correlator diverges at late times with some Lyapunov exponent $\lambda_L$:
\begin{equation}
\Tr[V(0) W(t) V(0) W(t) e^{-\beta H}] \sim e^{\lambda_L t} \;,  \qquad   W(t) = e^{\imath t H} W(0) e^{-\imath t H} \;.
\end{equation}

In  \cite{MalStaShe} Maldacena, Shenker and Stanford studied a four point correlator equally spaced on the thermal circle
$
F(t) = \Tr [yV(0) yW (t)yV(0) yW (t)]$ with, $y= {\cal Z}^{-1/4} e^{- \frac{\beta}{4} H } $
and found that under mild assumptions and for intermediate time scales\footnote{Comprised between the diffusion and scrambling times of the system. The diffusion time characterizes the response to a local perturbation; the scrambling time measures the spreading of quantum information across the system.} one gets:
\bee F(t) \simeq  a -\frac{b}{N^2} \;, e^{\lambda_L t} \; , 
\qquad t_d <  t < t_s  \; ,
\ee
where $N$ denotes the number of degrees of freedom of the system and the Lyapunov exponent $\lambda_L$ obeys an upper bound $\lambda_L  \le 2 \pi /\beta$, hereafter called the MSS bound.
Furthermore, they argued that saturation of this bound is a strong indication of the presence of a gravitational dual. 

This result generated an impressive volume of follow up work. 
In particular Kitaev \cite{Kit,Kit1} exhibited a one dimensional quantum mechanical model, reminiscent of a model previously studied by Sachdev and Ye \cite{SacYe}, which saturates the MSS bound and has a two dimensional gravitational dual. The action of the SYK model is:
\bee 
S(\psi) = \int dt  \biggl( \frac{\imath}{2} \sum_{i=1}^N \psi_i \frac{d}{dt}\psi_i - \imath^{q/2}	 \sum_{ 1 \le i_1 <  \cdots < i_q \le N}	J_{i_1 \cdots  i_q}	\psi_{i_1} \cdots \psi_{i_q}\biggr) \;,
\ee
with $J_I$ a quenched Gaussian random tensor coupling $\mathbb{E}( J_I J_{I'} ) = \delta_{II'}J^2 (q-1)! N^{-(q-1)}$ and $\psi_i$ an $N$-component Majorana Fermion. Somewhat surprisingly, this model is not only maximally chaotic, but it is also solvable in the $N \to \infty$ limit, two features which were previously considered incompatible. The SYK model becomes approximately reparametrization (i.e. conformally) invariant in the infrared, leading to a so called ``near $AdS_2$ / near $CFT_1$'' correspondence. Any nearly-extremal black hole should have a two dimensional ``throat" in which the radial distance to the horizon and the time should be the only relevant dimensions hence simple models of two dimensional gravitational theories dual to one dimensional $CFT$s should shed light on issues such as the information paradox.

The reason the SYK model can be solved is that, like for random tensors, its large $N$ limit is dominated by melonic graphs. 

\paragraph{Two Point Function.}
The first step in solving the model consists in determining its Euclidean two point function:
\bee
G(\tau) = \frac{1}{N}\sum_i\langle\psi_i(\tau)\psi_i(0) \rangle \;,\qquad
G(\omega) = \int_{-\infty}^\infty d{\tau}  \; e^{i\omega \tau}G(\tau) \; ,
\ee
and at finite temperature the Matsubara frequencies are quantized $\omega_n= \frac{2\pi }{\beta} (n + \frac{1}{2})$ and the Euclidean time is compact $-\beta/2\leq \tau\leq \beta/2$. 

\begin{figure}[htb]
\centerline{\includegraphics[width=1.3cm]{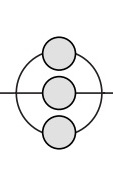}}
\caption{The self energy in the melonic limit.}
\label{fig : 2 point melonic}
\end{figure}

Denoting $\Sigma(\omega)$ the self-energy, that is the one particle irreducible amputated two-point function, the Schwinger-Dyson equation (SDE) of the model in the large $N$ limit reads:
\bee G^{-1} (\omega) = \imath \omega - \Sigma (\omega),\qquad  \Sigma (\tau) = J^2 G(\tau )^{q-1} \;,
\ee
The first equation is the usual SDE linking the dressed two-point function to the bare free part of the action $\imath \omega$ and the self-energy. The second one encodes the effect of taking the quenched average over the random couplings and states that at leading order in the large $N$ limit the self energy consists in a melon (i.e. two vertices connected by $q-1$ parallel edges) with dressed propagators, as depicted in Fig. \ref{fig : 2 point melonic}.

Taking the infrared limit, the free term in the SDE is suppressed and we are left with:
\bee 
\label{eq:convolution}
J^2 \int d \tau^\prime \;  G(\tau-\tau^\prime) G(\tau^\prime - \tau^{\prime \prime} )^{q-1} = -\delta (\tau - \tau^{\prime\prime}) \; ,
\ee
which is solved at zero and finite temperature respectively by:\footnote{
Eq.~\eqref{eq:convolution} is reparametrization invariant, that is if $G(\tau, \tau')$ is a solution then $[ f'(s) f'(s') ]^\Delta G(f(s),f(s'))$ is a also solution for any differentiable function $f$, and one maps the finite temperature equation to the zero temperature one via $f_\beta ( s ) = \tan (\pi s/\beta )$.}
\begin{equation}
 G (\tau)  \sim \vert \tau \vert^{-2 \Delta} \, {\rm sign}(\tau) \;,\qquad G ( s ) \sim | \sin (\pi s / \beta) |^{-2 \Delta} \ {\rm sign } ( s ) \;, 
\end{equation}
with $\Delta=1/q $ the infrared dimension of the field, for which the $q$-body interaction is marginal.

\paragraph{Four Point Function.} 
In order to check for chaotic behaviour, one needs to compute the four point function of the model:
 \begin{equation}
 \frac{1}{N^2}\sum_{1\leq i,j\leq N}
\langle  \psi_i(\tau_1)\psi_i(\tau_2)\psi_j( \tau_3)\psi_j(\tau_4)  \rangle =  G(\tau_{12})G(\tau_{34})\nonumber +\frac{1}{N} 
 \mathcal{F}(\tau_1,\tau_2 ; \tau_3, \tau_4)  \;,
 \end{equation}
where $\tau_{12} = \tau_1 - \tau_2$ and  $ \mathcal{F}(\tau_1,\tau_2 ; \tau_3, \tau_4)$ is the piece of the four point function which is connected in the channel $(12)-(34)$. 

The irreducible four point kernel \cite{Benedetti:2018goh} in the channel $(12)-(34)$ is the functional derivative of the self energy  with respect to the dressed propagator. It has the interpretation of a direct two particle to two particle scattering (that is without any intermediate  two particle state) and is therefore two particle irreducible in the channel $(12)-(34)$. The four point kernel amputated to the right is obtained by reestablishing two external dressed propagators:
\begin{equation}
K(\tau_1,\tau_2; \tau_3,\tau_4) = \int d\tau_1'd\tau_2'\;  \;
G(\tau_1,\tau_1') G(\tau_2 , \tau_2') \frac{\delta}{\delta G(\tau_1',\tau_2') } \Sigma(\tau_3,\tau_4)  \; .    
\end{equation}

Similarly to the SDE, one can write a Dyson equation for the four point function by making an analysis in two particle reducibility lines in the channel $(12)-(34)$. The four point function which transmits in the channel $(12)-(34)$ consists in direct propagation, one irreducible scattering $K$, two irreducible scatterings $KK$ and so on:
\begin{equation}\label{eq:resum}
\begin{split} 
& \mathcal{F} = \sum_{n \ge 0}  K^n {\cal  F}_0  \; , \qquad  
         \mathcal{F}_0 = -G(\tau_{13})G(\tau_{24}) + G(\tau_{14})G(\tau_{23})  \; , \crcr
& \mathcal{F}(\tau_1,\tau_2, \tau_3, \tau_4) = \int d \tau d \tau^\prime 
\left( \frac{1}{1-K} \right) (\tau_1,\tau_2 ; \tau, \tau^\prime)\mathcal{F}_0(\tau, \tau^\prime ; \tau_3, \tau_4) \; .         
\end{split} 
\end{equation}

\begin{figure}[htb]
\centering
\begin{subfigure}
\centering
\includegraphics[width=2cm]{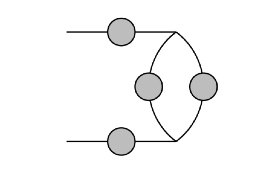}
\end{subfigure}
\hspace{1cm}
\begin{subfigure}
\centering 
\includegraphics[width=10cm]{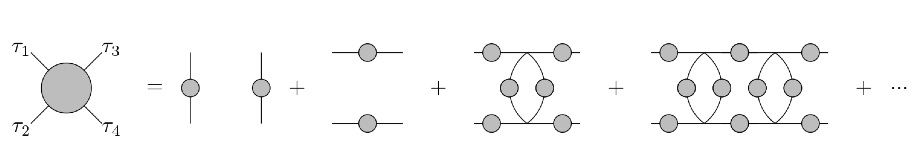}
\end{subfigure}
\caption{The kernel and the four point function at leading order.}\label{fig:allfourpoint}    
\end{figure}

At leading order in the large $N$ limit we have $\Sigma(\tau_3,\tau_4)  =J^2 G(\tau_3,\tau_4)^{q-1}$ hence the irreducible four point kernel reduces to one rung:
\begin{equation}
K(\tau_1,\tau_2 ; \tau_3, \tau_4)  = -J^2(q-1)G(\tau_{13})G(\tau_{24})G(\tau_{34})^{(q-2)} \;,
\end{equation}
as depicted in Fig.~\ref{fig:allfourpoint}.

In order to compute the four point function one needs to diagonalize the four point kernel. A complete basis in the space of two particle to two particle operators consists in the conformal  blocks $\Psi_h(\tau_1,\tau_2;\tau_3,\tau_4)$ corresponding to operators with dimensions in the principal series $h = \frac{1}{2} + \imath \mathbb{R}$ supplemented by discrete modes with $h$ even integers:
\begin{equation}
\begin{split}
& \frac{ - \delta(\tau_{13} ) 
\delta( \tau_{24} ) + \delta(\tau_{14})   
\delta( \tau_{23} ) }{2}  \crcr
& =  \int_{\frac{1}{2} -\imath \infty }^{\frac{1}{2} + \imath \infty } 
\frac{dh}{2\pi \imath} \; \rho(h)  \; \Psi_{h}(\tau_1,\tau_2 ; \tau_3,\tau_4)   +\sum_{n=1}^\infty \lim_{h\to 2n } \left[  (h-2n) \rho(h)  
\Psi_{h} (\tau_1,\tau_2 ; \tau_3,\tau_4) \right]\;,
\end{split} 
\end{equation}
where the density $\rho(h)$ has poles of order $1$ at $h=2n$ and $\rho(h) \Psi_h$ does not have any other poles in $h$. The conformal blocks diagonalize the four point kernel:
\begin{equation}
    \int d\tau d\tau' \; K(\tau_1,\tau_2; \tau , \tau') \, 
 \Psi_h(\tau,\tau'; \tau_2;\tau_4)
     = k(h)  \Psi_h(\tau_1,\tau_2; \tau_3;\tau_4) \; ,
\end{equation}
as well as the free term 
${\cal F}_0  \Psi_h = {\cal F}_0(h)\Psi_h  $.\footnote{It turns out that ${\cal F}_0(h)$ is in fact proportional to $k(h)$, but this is secondary here.}  Going back to the four point function in Eq.~\eqref{eq:resum} and inserting the resolution of the identity, we get:
\begin{equation}
    {\cal F} \sim \int_{C} \frac{dh}{2\pi \imath} \; \frac{1}{1-k(h)} \rho(h) {\cal F}_0(h) \Psi_h \;, 
\end{equation}
where the contour $C$ consists in the principal series $1/2 + \imath \mathbb{R}$ plus small circles around the poles at  $h=2n$. The integral is then evaluated by pushing the principal series to the right, cancelling the poles in $h=2n$ and picking up only the poles originating from the $\frac{1}{1-k(h)}  $ term, that is those values of $h$ such that $k(h) = 1$. This identifies the physical primary operators of the model and the integral is a sum over residues at the physical poles, giving the usual OPE expansion of the four point function.

There is however one caveat. It turns out that $k(2)=1$, meaning that one of the isolated contributions in the conformal blocks basis overlaps with a pole in the four point function. This mode needs to be desingularized and a careful analysis shows that it ends up being responsible for both the breaking of conformal invariance and the chaotic behaviour of the model. The second fact can be inferred by 
diagonalizing the ladder kernel in real time and solving its eigenvalue equation: only the mode corresponding to $h=2$ yields a solution, and its eigenfunction displays exponential growth \cite{MalStaYan}.

\paragraph{Tensor quantum mechanics.}
The slightly unpleasant feature of the SYK model is the quenched disorder average. Due to this average, the model is not a genuine quantum mechanical model, but a statistical ensemble. Witten \cite{witten2019syk} pointed out that one can eliminate the disorder average but keep 
the melonic large $N$ limit if one uses 
colored \cite{Gurau:2009tw,Gurau:2011xp}
tensors:
\bee 
S (\psi)= \int dt  \biggl( \frac{\imath}{2} \sum_{i} \psi_i \frac{d}{dt}\psi_i  -\imath^{q/2}	j 	\psi_{0} \psi_1 \cdots \psi_{D}\biggr) \;,
\ee
where $\psi$'s are $D+1$ real Fermionic tensors of order $D$. The pattern of index contractions in the interaction term is the one of the colored tensor model \cite{Gurau:2009tw}, incidentally the first tensor model for which a $1/N$ expansion has been established \cite{Gurau:2010ba,Gurau:2011aq,Gurau:2011xq,Gurau:2016lzk}. Following this remark, several other tensor quantum mechanical models have been proposed \cite{Klebanov:2018nfp,Pakrouski:2018jcc,Klebanov:2019jup,Gaitan:2020zbm}. Such models can be studied along the lines of the SYK model, and typically provide one dimensional conformal field theories in the infrared. 

The main difference between tensor quantum mechanical models and the SYK (or a matrix) model lies in the relation between the number of independent fields and the symmetry group.
In vector models, like the SYK, the number of field is $N$ and the dimension of the symmetry group is $N^2$. The only invariant one can build is the square of the field. For matrices one has $N^2$ fields and the symmetry group has dimension $N^2$. One can already build more invariants, the traces of the powers of the matrix. The situation is very different for tensors, for which one has $N^p$ fields, while the dimension of the symmetry group goes like $N^2$: one can build a plethora of invariants, namely all the trace invariants associated to $p$-colored graphs.

Although one needs to deal with the large number of invariants, tensor quantum mechanics has a fundamentally positive feature: contrary to the SKY model it fits into the paradigm of (one dimensional) local quantum field theory and one can study its generalizations to genuine tensor field theories in higher dimension.

\subsection{Tensor field theories}

After tensor quantum mechanics \cite{witten2019syk}, the next natural step is to consider tensor field theories in higher dimensions. Introduced by Klebanov and collaborators \cite{Giombi:2017dtl}, such models have been actively studied over the past several years \cite{Giombi:2017dtl,carrozza2016n,Bulycheva:2017ilt,Giombi:2018qgp, Choudhury:2017tax,Benedetti:2018goh,Klebanov:2018fzb,Kim:2019upg,
DeMelloKoch:2019tmo,Klebanov:2020kck,Peng:2016mxj,Jepsen:2023pzm}. The most common tensor field theory is built out of a (Bosonic) tensor field $\phi_{a^1a^2a^3}$ transforming in the three fundamental representation of the orthogonal group \cite{Giombi:2017dtl,carrozza2016n}.  Denoting triples of indices in bold $\pmb{a}=(a^1,a^2,a^3)$ its action writes (repeated indices are summed):
\begin{equation}
   S (\phi)= \int d^dx \bigg( \frac{1}{2} \phi_{\pmb{a} } (-\partial^2 + m^2)  \phi_{\pmb{a}} +
   \frac{1}{4} \big(\lambda_1
     P^1_{\pmb{a}\pmb{b}; \pmb{c}\pmb{d}}
 +  \lambda_2
     P^2_{\pmb{a}\pmb{b}; \pmb{c}\pmb{d}} + 
     \lambda \hat \delta^t_{\pmb{a}\pmb{b}\pmb{c}\pmb{d}} \big)
   \phi_{\pmb{a}} \phi_{\pmb{b}} \phi_{\pmb{c}} \phi_{\pmb{d}} \; ,
\end{equation}
where $-\partial^2 = -\partial_\mu\partial^\mu$ is the Laplacian operator in $d$ dimensions.
The three interaction terms are written in terms of the three invariant contraction patterns, double trace, pillow and tetrahedral:
\begin{equation}
\begin{split}
   &  \delta^{d}_{ \pmb{a}\pmb{b} ;\pmb{c}\pmb{d}} =\prod_{i=1}^3 \delta_{a^ib^i}\delta_{c^id^i} \; ,\qquad
 \delta^{p}_{ \pmb{a}\pmb{b} ; \pmb{c}\pmb{d}} = \frac{1}{3} \sum_{i=1}^3 \delta_{a^ic^i}\delta_{b^id^i} \prod_{j\neq i} \delta_{a^jb^j} \delta_{c^jd^j} \; ,
   \crcr
   & \delta^{t}_{ \pmb{a}\pmb{b}\pmb{c}\pmb{d}}= \delta_{a^1 b^1}\delta_{c^1d^2} \; \delta_{a^2c^2} \delta_{b^2d^2} \; 
   \delta_{a^3d^3} \delta_{b^3c^3} \; ,\crcr
&  P^1_{\pmb{a}\pmb{b}; \pmb{c}\pmb{d}} =
 \frac{3}{N^2}  \delta^{p}_{ \pmb{a}\pmb{b} ;\pmb{c}\pmb{d}} - \frac{3}{N^3}  \delta^{d}_{ \pmb{a}\pmb{b} ;\pmb{c}\pmb{d}} \;,\qquad 
  P^2_{\pmb{a}\pmb{b}; \pmb{c}\pmb{d}}  = \frac{1}{N^3}  \delta^{d}_{ \pmb{a}\pmb{b} ;\pmb{c}\pmb{d}} 
  \;  , \qquad 
  \hat \delta^t_{\pmb{a}\pmb{b}\pmb{c}\pmb{d}} = \frac{1}{N^{3/2}} \delta^t_{\pmb{a}\pmb{b}\pmb{c}\pmb{d}}\;. 
\end{split}
\end{equation}

Introducing an infrared scale $\mu$, the renormalization group flow of the model is encoded in anomalous field dimension and the beta functions, which describe the variation of the effective couplings with the infrared scale.
Rescaling the couplings as $\tilde g = g/(4\pi)^2$ etc. and correspondingly the beta functions $\tilde\beta =\mu\partial_\mu \tilde g$, we obtain at two loops in $4-\epsilon$ dimensions \cite{Giombi:2017dtl}:
\begin{equation}\label{eq:betain4}
\tilde \beta = -\epsilon \tilde g + 2 \tilde g^3 \;, \quad
\tilde \beta_1 = -\epsilon \tilde g_1 + 2 (\tilde g_1^2 +\tilde g^2) -2\tilde g^2 \tilde g_1 
\;,\quad \tilde \beta_2 = -\epsilon \tilde g_2 + 2 (\tilde g_2^2 +3\tilde g^2) -5\tilde g^2 \tilde g_2 \;,
\end{equation}
and the anomalous field dimension $\eta_\phi= \tilde g^2 /2 $. This system of beta functions admits a fixed point $g\sim \sqrt{\epsilon}$ and $g_{1,2} \sim \imath \sqrt{\epsilon}$ corresponding to a conformal field theory, but the latter is unstable due to the presence of an operator with dimensions $2+\imath \alpha$ \cite{Giombi:2017dtl,Benedetti:2021qyk}

\paragraph{The long range $O(N)^3$ model.} A more interesting infrared behaviour is obtained in a long range version of the model \cite{Benedetti:2019eyl,Benedetti:2018ghn,Gurau:2019qag,Benedetti:2019ikb,Benedetti:2021wzt}:
\begin{equation}
   S(\phi) = \int d^dx \bigg( \frac{1}{2} \phi_{\pmb{a} } \big( (-\partial^2)^\zeta + m^{2\zeta} \big)  \phi_{\pmb{a}} +
   \frac{1}{4} \big(\lambda_1
     P^1_{\pmb{a}\pmb{b}; \pmb{c}\pmb{d}}
 +  \lambda_2
     P^2_{\pmb{a}\pmb{b}; \pmb{c}\pmb{d}} + 
    \imath \lambda \hat \delta^t_{\pmb{a}\pmb{b}\pmb{c}\pmb{d}} \big)
   \phi_{\pmb{a}} \phi_{\pmb{b}} \phi_{\pmb{c}} \phi_{\pmb{d}} \; ,
\end{equation}
with $\zeta<1$ a non trivial scaling exponent. 
Note that we have explicitly factored an $\imath$ imaginary unit in the tetrahedral coupling constant.
The dimension of the field respects $2\Delta_\phi+2\zeta=d$ and the quartic interaction is marginal for $4\Delta_\phi = d$, that is $\zeta = d/4$. The ultraviolet divergences are regularized by setting $\zeta=(d+\epsilon)/4$, and the infrared ones by some regulator at the infrared scale $\mu$, for instance by shifting $ (-\partial^2)^\zeta = (-\partial^2 + \mu^2 )^\zeta$.  Due to the long range propagator, the model \emph{does not exhibit} a wave function renormalization and the dimension of the field is fixed along the renormalization group flow. 

Denoting $\tilde g (4\pi)^{d/2}\Gamma(\zeta)^2 = g$,
the beta functions at one loop are:
\begin{equation}
\tilde \beta = -\epsilon \tilde g \;,\quad 
\tilde \beta_1 = -\epsilon \tilde g_1 + 2 \frac{\Gamma(d/4)^2}{ \Gamma(d/2) }( \tilde g_1^2 - \tilde g^2) \;,\quad
\tilde \beta_2 = -\epsilon \tilde g_2 + 2 \frac{\Gamma(d/4)^2}{ \Gamma(d/2) }( \tilde g_2^2 - 3 \tilde g^2)  \;.
\end{equation}
The model displays four lines of fixed points in the critical case $\epsilon =0$, and their stability can be inferred from the eigenvalues of the stability matrix $\partial{\tilde g_a} \tilde \beta_b$:
\begin{equation}
\tilde g =\tilde g^\star \; ,  \; \; \tilde g^\star_1 =\pm \tilde g^\star \; , \; \; \tilde g^\star_2 = \pm\sqrt{3} \tilde g^\star  \;,\qquad
 \delta h = 0 \;, \; \; 
 \delta h_1 = 4 \frac{\Gamma(d/4)^2}{\Gamma(d/2)} \tilde g_1^\star \; ,\; \; 
 \delta h_2 = 4 \frac{\Gamma(d/4)^2}{\Gamma(d/2)} \tilde g_2^\star \; .
\end{equation}

\paragraph{Scaling operators.} In order to study the fixed point theory one needs first to identify the scaling operators at the fixed point. Let us consider the long range free action 
perturbed by a general perturbation $S(\phi) =S_0(\phi) +S^{\rm int}(\phi)$ with:
\begin{equation}
S_0 (\phi) = \int d^dx \; \phi(x) ((-\partial^2)^\zeta + \mu^{2\zeta}) \phi(x) \;,\qquad  
S^{\rm int} (\phi) =\sum_{a}\lambda_a \int d^d x \; O_a(x) \;, 
\end{equation}
where we included an infrared regulator in the free action and the interaction is 
a sum over integrated local operators which we denote schematically $O_a(x) = (\partial^{p_a} \phi^{n_a} ) (x)$. 

The fundamental field writes in terms of its dimensionless 
version\footnote{This 
is the physicist way to account for the transformation of $\phi$ under dilatation, $\phi(\Omega x) = \Omega^{-\Delta_\phi} \phi(x) $.}
as $\phi(x) = \mu^{\Delta_\phi} \phi'(\mu x)$,
hence $O_a(x) = \mu^{\Delta_a} O'_a(\mu x) $ where $\Delta_a = n_a\Delta_\phi + p_a$ is the 
canonical dimension of $O_a$ and $O'_a$ is the dimensionless version of $O_a$. The local operators $O_a$ are eigenoperators of the dilatation operator $x \cdot \partial_x = x^\nu\partial_{x^\nu}$:
\begin{equation}
  0 =  \mu\frac{d}{d\mu} O_a(x) = \mu\frac{d}{d\mu} \bigg( \mu^{\Delta_a}O_a'(\mu x) \bigg) = 
  (\Delta+ x\cdot \partial_x) O_a(x) \; .
\end{equation}

On dimensional grounds the connected correlations of the fundamental field write as:
\begin{equation}
\begin{split}
 \langle \phi (x_1) \dots \phi (x_n) \rangle_c & = \int_c [d\phi] \;e^{-S_0(\phi) - \sum_a \lambda_a \int d^dx \; O_a(x) } \phi (x_1) \dots \phi (x_n) \crcr
&  =\mu^{n \Delta_\phi} B ( \mu^{-d+\Delta_a}\lambda_a ; \mu x_i )     \;,
\end{split}
\end{equation}
where the subscript $c$ signifies that one takes only the connected contribution to the correlation and the bare expansion $B$ exhibits poles in $1/\epsilon$. We substitute the bare couplings in terms of dimensionless renormalized couplings $g$ and counterterms 
$\lambda_a = \mu^{d-\Delta_a} [ g_a + \sum_{p\ge 1} \epsilon^{-p}C^{(p)}_a(g) ]$, and fix the counterterms by requiring that the poles in $1/\epsilon$ cancel, that is:
\begin{equation}
\begin{split}
\langle \phi (x_1) \dots \phi (x_n) \rangle_c & =
 \int_c [d\phi] \;e^{-S_0(\phi) - \sum_a  [ g_a + \sum_{p\ge 1} \epsilon^{-p}C^{(p)}_a(g) ] \mu^{d-\Delta_a} \int d^dx \; O_a(x) } \phi (x_1) \dots \phi (x_n)\crcr 
& = \mu^{ n\Delta_\phi } R ( g_a ; \mu x_i )     \;,     
\end{split}
\end{equation}
where the renormalized expansion $R$ has no poles in $1/\epsilon$. Note that here we use the fact that, due to the long range propagator, there is no anomalous scaling of the fundamental field $\phi$.

The dimensionless renoramlized operators are\footnote{This definition of the renormalized operators would need to be revisited if the field had anomalous scaling.}
$\int d^d x [O_a](x) = \mu^{-(d-\Delta_a)} \frac{ \partial S^{\rm int} }{\partial g_a}$ and can be written in terms of the bare operators and the dimensionless mixing matrix:
\begin{equation}
    [O_a ](x) = \sum_b \mu^{\Delta_a} Z_{a;b}\mu^{-\Delta_b } O_b(x) \;, \qquad Z_{a;b} = 
   \mu^{-(d-\Delta_b)} \frac{\partial \lambda_b}{\partial g_a} \; .
\end{equation}
The renormalized operators are central in a renormalization group analysis, as the insertion of such operators into correlation functions expressed in the renormalized expansion does not generate poles in $1/\epsilon$:
\begin{align}
- \partial_{g_a}  \langle  \phi (x_1) \dots \phi (x_n) \rangle_c  =
     \mu^{d-\Delta_a} \left\langle  \int d^dx [O_a] (x) \;  \phi (x_1) \dots \phi (x_n)  \right\rangle_c
\; .
\end{align}

The beta functions $\mu\partial_{\mu}g$ encode the change of the renormalized couplings with the infrared scale at fixed bare couplings and also write in terms of the mixing matrix. Using matrix notation and denoting $\Delta = \Delta_a \delta_{ab}$ the diagonal matrix of operator dimensions we have: 
\begin{equation}\label{eq:beta}
\mu \frac{d}{d\mu} \lambda  = 0 \Rightarrow
 \lambda (d-\Delta) +  \beta   Z \mu^{d-\Delta}= 0 \Rightarrow \qquad \beta = -\lambda(d-\Delta) \mu^{\Delta - d} Z^{-1} \; ,
\end{equation}
where one needs to substitute $\lambda$ and $Z$ in terms of the renormalized couplings $g$: the scale $\mu$ drops out and the $\beta$ functions do not exhibit poles in $1/\epsilon$.\footnote{Note that at a fixed point the beta functions are 0, that is $Z^{-1}$ has a non trivial (left) kernel and the mixing matrix is not invertible.}

Like the bare couplings, which are such that $\mu \frac{d}{d\mu}\lambda_a = 0$, the bare operators are such that $\mu\frac{d}{d\mu} O_a= 0$, hence $\mu\frac{d}{d\mu} S^{\rm int} (\phi) =0$. Expressing $S^{\rm int}(\phi)$ in terms of renormalized couplings and observing that, acting on the renormalized quantities, $\mu \frac{d}{d\mu} = \mu\partial_\mu + \beta_b \partial_{g_b}$ we have:
 \begin{equation}
 \begin{split}
 \partial_{g_a} \left( \mu \frac{d}{d\mu} S^{\rm int} (\phi) \right)=0
& = (\mu \partial_\mu \partial g_a + (\partial_{g_a} \beta_b)\partial_{g_b} + 
  \beta_b \partial_{g_a} \partial_{g_b}  
  ) S^{\rm int} (\phi)  \crcr
&  = \int d^dx \bigg\{ (\mu \partial_\mu \delta_{ab}+ \partial_{g_a} \beta_b) ( -  \mu^{d-\Delta_b} [O_b](x)   ) \bigg\} + \beta_b \partial_{g_a} \partial_{b_b} S^{\rm int} (\phi) \;.
 \end{split}
 \end{equation}
  
At a fixed point $\beta =0$ and the last term drops out. Denoting $Y_{ab} = \partial_{g_a} \beta_b$ the stability matrix, we have:
\begin{equation} \label{eq:firstqe}
  (  \mu \partial_\mu +d +Y ) \mu^{-\Delta}  [O](x)  =0  \; .
\end{equation}
The $\mu\partial_\mu$ term captures the behaviour of the renormalized operator $[O]$ under dilatation:
\begin{equation}\label{eq:secondeq}
  \mu \partial_\mu \bigg( 
\mu^{ -\Delta}  [O](x)  \bigg) = 
\mu \partial_\mu ([O]'(\mu x))   = \mu^{-\Delta}(x\cdot \partial_x) [O](x) \;,
\end{equation}
hence, according to \eqref{eq:firstqe}, the action of the dilatation on $[O]$ is not diagonal. The scaling operators which diagonalize the action of the dilatation at the fixed point\footnote{As the $\beta$ functions are zero, at the fixed point $ \mu \frac{d}{d\mu} = \mu \partial_\mu + \beta_b \partial_{g_b} = \mu \partial_\mu  $. } $\mu\frac{d}{d\mu} \{O_a\}= \mu \partial_\mu \{O_a\} =0$ are the eigen operators of the stability matrix.
Indeed, diagonalizing $Y = U\nu U^{-1}$. we have:
\begin{equation}
\begin{split}
& \{O\}(x) = \mu^{d+\nu} U^{-1} \mu^{ -\Delta} [O](x) \;, 
\crcr
& \mu \frac{d}{d\mu} \{ O\} (x)=  
( d+\nu  ) \{O \}(x) + \mu^{d+\nu} U^{-1} 
\mu \partial_\mu \bigg( 
\mu^{ -\Delta}  [O](x)  \bigg) =0 \; ,
\end{split}
\end{equation}
where we used Eq.~\eqref{eq:firstqe}. At the same time, Eq.~\eqref{eq:secondeq} implies $\mu \frac{d}{d\mu}\{O\}(x) =  ( d+\nu +  x  \cdot \partial_{ x} ) \{ O\} (x)$, signifying that $\{O\}$ is an eigen operator of the dilatation with eigenvalue $d+\nu$.\footnote{Deriving Eq.~\eqref{eq:beta} we relate the stability matrix with the matrix of anomalous dimensions 
$\gamma =- \mu ( \frac{d}{d\mu}Z) Z^{-1}=- \beta ( \partial_g Z ) Z^{-1}$ as $d+Y = Z\Delta Z^{-1} + \gamma$.
Note that, although $\beta =0$ at a fixed point, $\gamma = - \beta ( \partial_g Z ) Z^{-1}$ is not zero, as the mixing matrix is singular.}

 One can analyze the scaling operators in perturbation theory. Besides the quartic operators, one can for instance compute the anomalous scaling dimension of operators of the form $\phi(-\partial^2)^n\phi$. Denoting $\tilde g_{(n)}$ the couplings of $\phi(-\partial^2)^n\phi$ and $\Delta_\phi^\star = d/4$ the dimension of the field in the critical case, we get at lowest orders in perturbation theory:
\begin{equation}
\begin{split}
\tilde \beta_{(1)} &= -(d-2\Delta_\phi^*) \tilde g_{(1)} +  2 \frac{\Gamma(d/4)^2}{\Gamma(d/2)} \tilde g_{(1)} \tilde g_2 \;, \crcr
\tilde \beta_{(n)} &= -(d-2\Delta_\phi^* - 2n ) \tilde g_{(n)} + 6 \frac{\Gamma(d/4)^4}{\Gamma(d/2)}  \; \frac{\Gamma(n+1-d/2)}{n\,\Gamma(n+d/2)\Gamma(1-d/2)}   \tilde g_{(n)} \tilde g^2 \; ,
\end{split}
\end{equation}
where at first order we only take into account the 
back reaction of adding an insertion of the operator $\phi(-\partial^2)^n\phi$ in a two point function.  

\paragraph{Conformal limit.}
Although the model does not have a local energy-momentum tensor, the fixed point theories are conformally invariant \cite{Benedetti:2020yvb}.
One can then use non perturbative conformal partial waves techniques
\cite{Benedetti:2021wzt,Benedetti:2019ikb} and identify the primaries of the theory corresponding to the eigenvalue $k(h,J)=1$ of the four point kernel:
\begin{equation}
\begin{split}
&\int dy\; dz \; K(x_1,x_2 ; y,z ) \langle \phi(y) \phi(z) O_{h,J}(x_3)  \rangle 
= k(h,J) \;  \langle \phi(x_1) \phi(x_2) O_{h,J}(x_3)  \rangle \; , \crcr
& k(h,J) = 3g^2 \Gamma\left(\frac{d}{4} \right)^4 \; 
\frac{\Gamma\left( -\frac{d}{4} + \frac{h+J}{2} \right) \Gamma\left( \frac{d}{4} - \frac{h-J}{2} \right) }{\Gamma\left( \frac{3d}{4} - \frac{h-J}{2} \right)  \Gamma\left( \frac{d}{4} + \frac{h+J}{2} \right) }  \; ,
\end{split}
\end{equation}
where the solutions take the from $h = \frac{d}{2} + 2n + J + \delta h_{(n)}$. The perturbative and conformal partial waves computations agree and one concludes that only the bilinear primaries  contribute to the OPE of four fundamental fields in the large $N$ limit.

One can push the conformal partial waves techniques much further, not only to derive the OPE coefficients, but to use them in order to resum infinite series of diagrams
\cite{Benedetti:2021wzt,Benedetti:2019ikb}.
One can thus check for instance that the sphere free energy of the model decreases between ultraviolet and infrared renormalization group fixed points, that is the model respects the $F$-theorem.

\paragraph{Asymptotic freedom in a Bosonic field theory.}

A tantalizing fact \cite{Berges:2023rqa} is that the beta functions in \eqref{eq:betain4} in dimension $d=4$ describe an \emph{asymptotically  free} theory for an imaginary tetrahedral coupling. Indeed, as the corresponding beta function is cubic, one can flip its sign by going to a purely imaginary coupling.  The renormalization group flow is controlled by the tetrahedral coupling which goes to zero in the ultraviolet and diverges  in the infrared. The other two couplings remain real along the flow, and one identifies a trajectory along which they are always positive. This ensures that, although the action acquires an imaginary part, the theory is well behaved as the real part of the action is bounded from below. The study of the resulting asymptotically free Bosonic theory in 4 dimensions is ongoing.

\section{Conclusion and Perspective}

We have briefly reviewed to current applications of random tensors to several flavours of quantum field theory. However, random tensors are expected to play an increasingly important role in many other areas of mathematics, physics, and even computer science.

In this context, a very important question is to identify interesting invariant quantities that characterize random tensors. For matrices one has eigenvalues, and the large $N$ distribution of 
the eigenvalues holds crucial information about the underlying random matrix. What becomes of this for tensors? Tensors also have eigenvalues, but they are much more complicated than the ones of matrices. For one, there are several notions of eigenvalues for tensors: the most useful one is that the eigenvalues and eigenvectors are the critical points of the function $\sum_{a}T_{a^1\dots a^D} v^{a^1} \dots v^{a^D}$ subjected to the constraint $\sum_a v^{a} v^a =1$. However, there are two problems to reckon with: first, the number of eigenvalues is exponentially large in the size of the tensor. Second, there is no restriction of the tensors (like real symmetric for instance) which generates only real eigenvalues: one always finds both real and complex eigenvalues and eigenvectors. One can perform various countings of the real eigenvalues \cite{Sasakura:2022zwc,Sasakura:2022iqd,Sasakura:2022axo}, but what to do with the complex ones? This question still awaits a satisfactory answer. 

An cheap alternative is, instead of focusing of the tensor eigenvalues, to focus on the eigenvalues (or rather the spectrum) of a matrix built out of the tensor. One such example is the tensor resolvent proposed in \cite{Gurau:2020ehg}. While the resolvent has the advantage that, being an operator, it has a well defined spectrum, the exact relation between its spectrum and the tensor eigenvalues is not trivial.

\section*{Acknowledgements}

R.G. is supported by the Deutsche Forschungsgemeinschaft (DFG, German Research Foundation) under Germany’s Excellence Strategy EXC2181/1-390900948 (the Heidelberg STRUCTURES Excellence Cluster) and the European Research Council (ERC) under the European Union’s Horizon 2020 research and innovation program (grant agreement No818066).

\end{document}